\documentclass[aps,twocolumn,pra,superscript,floatfix,superscriptaddress,showpacs]{revtex4-1}

\usepackage{graphicx}
\usepackage{dcolumn}
\usepackage{bm}
\usepackage{color,soul}

\usepackage{amsmath}

\usepackage[colorlinks=true,%
            linkcolor=blue,%
            urlcolor=blue,%
            citecolor=blue,%
            filecolor=blue,%
            bookmarksopen=true,%
            pdfauthor={AGarrido},%
            pdftitle={QDinterfometerMZM},%
            pdfsubject={QDinterfometerMZM_Manuscript},%
            pdfpagemode=UseOutlines]{hyperref}

\definecolor{applegreen}{rgb}{0.55, 0.71, 0.0}
\definecolor{antiquefuchsia}{rgb}{0.57, 0.36, 0.51}
\definecolor{amethyst}{rgb}{0.6, 0.4, 0.8}

\begin{document}


\title{Thermoelectric transport through a Majorana zero modes interferometer}

\author{A. P. Garrido}
\email{alejandro.garridoh@usm.cl}
\affiliation{ 
Departamento de F\'isica, Universidad T\'ecnica Federico Santa Mar\'ia, Av. Espa\~na 1680, Casilla 110V, Valparaiso, Chile.
}%

\author{D. Zambrano}
 \affiliation{ 
Departamento de F\'isica, Universidad T\'ecnica Federico Santa Mar\'ia, Av. Espa\~na 1680, Casilla 110V, Valparaiso, Chile.
}%
\author{J. P. Ramos-Andrade}
\affiliation{Departamento de F\'isica, Universidad de Antofagasta, Av. Angamos 601, Casilla 170, Antofagasta, Chile.}
\author{P. A. Orellana}%
 \affiliation{ 
Departamento de F\'isica, Universidad T\'ecnica Federico Santa Mar\'ia, Av. Espa\~na 1680, Casilla 110V, Valparaiso, Chile.
}%

\date{\today}

\begin{abstract}
In this study, we examine the thermoelectric characteristics of a system consisting of two topological superconducting nanowires, each exhibiting Majorana zero modes at their ends, connected to leads within an interferometer configuration. By employing Green's function formalism, we derive the spectral properties and transport coefficients. Our findings indicate that bound states in the continuum (BICs) manifest in symmetric setups, influenced by the length of the wires and coupling parameters. Deviations of the magnetic flux from specific values transform BICs into quasi-BICs with finite width, resulting in conductance antiresonances. The existence and interplay of Majorana zero modes enhance thermoelectric performance in asymmetric configurations. Modulating the magnetic flux transitions BICs into quasi-BICs significantly enhances the Seebeck coefficient and figure of merit, thereby proposing a strategy for optimizing thermoelectric efficiency in systems based on Majorana zero modes.

\end{abstract}

\maketitle

\section{\label{sec:intro}Introduction}

Recently, topological superconductor nanowires (TSCNs) have attracted significant attention in condensed matter physics for their potential in quantum computing \cite{kitaev2003fault, nayak2008non, pachos2012introduction, beenakker2013search, laflamme2014publisher, albrecht2016exponential}. Exotic fermionic quasiparticles predicted within this framework, being their own antiquasiparticles, have emerged \cite{majorana1937teoria, wilczek2009majorana, franz2010race}. These quasiparticles, known as Majorana zero modes (MZMs), are localized in topological superconductors. The MZMs exhibit non-Abelian statistics and are manipulated through braiding operations \cite{kraus2013majorana, alicea2011non}, making them ideal for fault-tolerant quantum computation \cite{kitaev2001unpaired, bravyi2002fermionic, kitaev2003fault, nayak2008non, leijnse2011quantum, pachos2012introduction, kraus2013majorana, albrecht2016exponential}. They are predicted at the ends of a TSCN comprising a semiconductor-superconductor nanowire with strong spin-orbit interaction under a magnetic field.
The aforementioned system can be viewed as a realization of a Kitaev chain \cite{kitaev2001unpaired, kitaev2003fault, moore2009next}, where the coupling between the two MZMs located at opposite ends of the wire decays exponentially with the wire's length \cite{albrecht2016exponential}. This enables the construction of a qubit that is topologically protected from decoherence by local perturbations \cite{wu2012tunneling, kitaev2001unpaired, kraus2013majorana, albrecht2016exponential, semenoff2006teleportation, tewari2008testable}. Mourik and collaborators achieved the first physical realization of this system, reporting zero-bias anomalies in conductance as evidence of the presence of MZMs \cite{mourik2012signatures}. However, these anomalies do not always provide conclusive evidence of MZMs, highlighting the need for custom-designed experimental protocols \cite{deng2012anomalous, mourik2012signatures, das2012zero, lee2012zero, finck2013anomalous, churchill2013superconductor, zambrano2018bound, ramos2019fano}. 
An alternative method to identify MZMs involves using thermoelectric measurements, which offer distinct advantages by exploring their unique transport signatures. Conventional thermoelectric measurement techniques, developed in the early 1990s \cite{Molenkamp1992:PRL,HvanHouten_1992:SST}, have developed into potent tools for detecting chargeless MZMs. These techniques can disclose MZM signatures through thermal conductance \cite{Fu2008:PRL, Bauer2021:PRB}, voltage thermopower \cite{Dolgirev2019:RRL,Hou2013:PRB,Sela2019:PRL}, or the breach of the Wiedemann-Franz (WF) law \cite{Giuliano2022:PRB,Buccheri2022:PRB,Benjamin2024:EL}, providing complementary evidence beyond zero-bias anomalies.

Contemporary protocols facilitate the identification of topological phases featuring MZMs in superconductor-semiconductor devices \cite{pikulin2021protocol}. This is achieved through a three-terminal configuration comprising two normal leads alongside a superconducting lead, which employs non-local conductance measurements to observe topological transitions via variations in the energy gap. The experiments carried out by Aghaee et al. substantiated these findings in heterostructures, affirming thus the presence of topological superconductivity and MZMs \cite{aghaee2023inas}. Recent advances in quantum computing exploit the principles of Majorana physics; notably, the Majorana-1 processor incorporates MZMs, resulting in enhanced scalability \cite{aasen2025roadmap}.
Conversely, bound states in the continuum (BICs) remain stable even when their energy levels reside within the domain of continuum states \cite{hsu2016bound}. Originally predicted by von Neumann and Wigner \cite{vonNeumann-Wigner}, BICs have garnered considerable interest, especially within photonic systems. Furthermore, owing to the similar interference phenomena observed in both electronic and photonic systems, the potential presence of BICs in electronic systems has been posited \cite{ramos2014bound, hsu2016bound, zambrano2018bound, grez2022bound, garrido2023bound}.

In this work, we study a system composed of two normal leads that interact in parallel with two TSCNs that host MZMs at their ends, forming an interferometer configuration, as illustrated in FIG.~\ref{fig1}. Our primary focus is on the thermal and electrical conductances between the normal leads and the spectral functions of the MZMs, computed using the Green function (GF) formalism. By modulating the magnetic flux within the interferometer, we discern signatures of quantum interference phenomena and the interaction between MZMs and BICs. Our findings indicate that BICs manifest in high-symmetry configurations, depending on the coupling strength between the TSCNs and the leads, as well as the lengths of the TSCNs. Moreover, we detect suppression in electronic and thermal conductance as a function of external magnetic flux, occurring within the mentioned symmetric configurations. We ascertain that the interaction between MZMs and BICs can be either triggered or inhibited by the magnetic flux, demonstrating the potential of this external parameter to effectively control these states. Finally, the annihilation of BICs by magnetic flux and/or asymmetry in couplings can enhance the response in thermopower and figure of merit and then enhance the thermoelectric efficiency of the system.

The structure of this paper is organized as follows. Section \ref{secII} elucidates the model along with the methodology used to derive the quantities of interest; Section\ \ref{secIII} presents the results and their subsequent discussion; and Section\ \ref{secIV} offers the concluding remarks.

\section{Model and method}\label{secII}

\begin{figure}[t]
    \centering
    \includegraphics[width=1.00\linewidth]{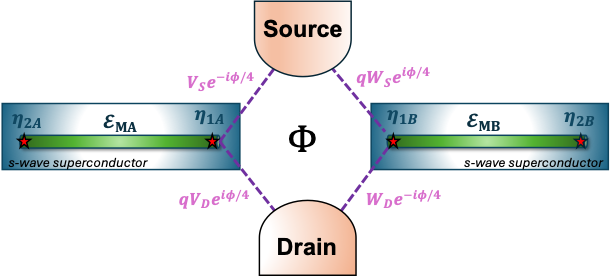}
    \caption{
    Schematic representation of the system under study: The TSCN $A(B)$ is connected to the lead S(D) and hosts two MZMs, $\eta_{_{1,A(B)}}$ and $\eta_{_{2,A(B)}}$, at its ends, with the intracoupling Majorana term given by $\varepsilon_{MA(B)}$. An external magnetic flux $\Phi$ threading the interferometer is considered. The parameters $V_{S(D)}(W_{S(D)})$ represent the couplings among the system's components, while $q$ is a dimensionless asymmetry parameter, where $q=1$ ($q=0$) corresponds to a closed (open) system, as explained later in the main text.}
    \label{fig1}
\end{figure}

We consider an interferometer configuration in which each TSCN is connected to two normal leads, S and D, and hosts MZMs at both ends, as schematically shown in Fig.\ \ref{fig1}. We model the system using an effective low-energy Hamiltonian of the following form:

\begin{equation}
    H = H_{\text{leads}}+H_{\text{M}}+H_{\text{M-leads}}\,,\label{eq1}
\end{equation}
where the first term on the right-hand side corresponds to the regular electronic contribution of the leads, given by  
\begin{align}
    H_{\text{leads}}      &= \sum_{\alpha,{\bf k}}\varepsilon_{\alpha,{\bf k}}a^{\dagger}_{\alpha,{\bf k}}a_{\alpha,{\bf k}}\,,\label{2}
\end{align}
where the operator $a^{\dagger}_{\alpha,{\bf k}}(a_{\alpha,{\bf k}})$ is the electron creation (annihilation) operator with momentum \textbf{k} and energy $\varepsilon_{\alpha,{\bf k}}$ in lead $\alpha=\text{S,D}$.  

The middle and last terms in the Hamiltonian presented in Eq.\ (\ref{eq1}) correspond to MZM-related terms, specifically MZM$_{\beta}$–MZM$_{\beta}$ and TSCN$_\beta$–lead$_\alpha$ couplings, given by

\begin{equation}
    H_{\text{M}} = \sum_{\beta}i\varepsilon_{_{M,\beta}}\eta_{_{1,\beta}}\eta_{_{2,\beta}}\label{3}\,,
\end{equation}
\begin{align}
    H_{\text{M-leads}}      &= \sum_{\beta}\left\{\sum_{\alpha,{\bf k}}(t_{\alpha,{\bf k}, \beta}a_{\alpha,{\bf k}} - t^*_{\alpha,{\bf k}, \beta}a^{\dagger}_{\alpha,{\bf k}})\eta_{_{1,\beta}}\right\}\,,\label{4}
\end{align}
where $\eta_{_{j,\beta}}$ denotes the MZM operator (with $j=1,2$ and $\beta=A,B$) and satisfies both $\eta_{_{j,\beta}}=\left[\eta_{_{j,\beta}}\right]^{\dagger}$ and $\{\eta_{_{j,\beta}},\eta_{_{j',\beta'}}\}=\delta_{j,j'}\delta_{\beta,\beta'}$. Additionally, $\varepsilon_{M,\beta}\propto \exp{(-L_{\beta}/\zeta)}$ represents the coupling amplitude between two MZMs in the same TSCN, where $L_{\beta}$ corresponds to the wire's length and $\zeta$ denotes the superconducting coherence length.  
The parameter $t_{\alpha, {\bf k},\beta}(\varphi_{\alpha, \beta})=t_{\alpha,{\bf k},\beta}^{(0)}\exp{[i\varphi_{\alpha,\beta}]}$ describes the TSCN$_\beta$–lead$_\alpha$ tunnel matrix element, where an Aharonov-Bohm (AB) phase is included to model the magnetic flux $\Phi$ across the interferometer. We adopt a symmetric gauge such that $\varphi_{_{D,A}}=-\varphi_{_{D,B}}=-\varphi_{_{S,A}}=\varphi_{_{S,B}}=\phi/4$, with $\phi=2\pi\Phi/\Phi_{0}$ and $\Phi_{0}=h/e$ being the quantum flux, where $h$ is Planck’s constant and $e$ the electron charge.

The GF is obtained from $(\mathbf{G}^{r})^{-1}= \mathbf{g}_0^{-1} + i\pi\mathbf{\Upsilon}\mathbf{\Upsilon}^\dagger$, where $\mathbf{g}_0^{-1}$ represents the Green’s function of the isolated MZMs. The matrix $\mathbf{\Upsilon}$ describes the coupling between the scatterer $(H_{\text{M}})$ and the leads. Since only MZM-1$\beta$ is coupled to the leads, the matrix representation of $\mathbf{g}_0^{-1}$ and $\mathbf{\Upsilon}$ can be expressed in the basis $\{\eta_{_{1A}}, \eta_{_{1B}}\}$, given by

\begin{equation}
    \mathbf{g}_0^{-1}= \left( 
    \begin{matrix}
        \omega      & -i\varepsilon_{_{MA}} & 0 & 0 \\
        i\varepsilon^*_{_{MA}} & \omega & 0                    & 0 \\
        0 & 0 & \omega                 & -i\varepsilon_{_{MB}} \\
        0 & 0 & i\varepsilon^*_{_{MB}} & \omega 
    \end{matrix}
    \right)\,,
    \label{g_0r}
\end{equation}

and
\begin{equation}
    \mathbf{\Upsilon}= \left( 
    \begin{matrix}
        -V_S e^{-i\phi/4} & -qV_D e^{i\phi/4} & V_S e^{i\phi/4}& qV_D e^{-i\phi/4} \\
        0& 0 & 0 & 0 \\
        -qW_S e^{i\phi/4}& -W_D e^{-i\phi/4}& qW_S e^{-i\phi/4} & W_D e^{i\phi/4} \\
        0 & 0 & 0 & 0 
    \end{matrix}
    \right)\,.
    \label{Gr}
\end{equation}

The Hamiltonian described in Eq.(\ref{eq1}) is spinless since only electrons with one spin projection will couple to the MZMs  \cite{ruiz2015interaction}.

The transmission probability is calculated from the expression

\begin{equation}
    \mathcal{T}(\omega) = \text{Tr}\{\mathbf{G}^{a}(\omega)\mathbf{\Gamma}^{D}\mathbf{G}^{r}(\omega)\mathbf{\Gamma}^{S}\}\label{11} \,,  
\end{equation}
where $\mathbf{G}^{r(a)}(\omega)$ is the system retarded (advanced) GF in the energy domain, and is obtained from

\begin{widetext}
\begin{equation}
    \left[\mathbf{G}^{r}\right]^{-1}= \left( 
    \begin{matrix}
        \omega + i\sum_\alpha \Gamma_{A}^{\alpha}  & -i\varepsilon_{_{MA}} & K  & 0\\ 
        
        i\varepsilon_{_{MA}}^{*} & \omega  & 0 & 0 \\
        
        K  & 0 & \omega + i\sum_\alpha \Gamma_B^{\alpha}  & -i\varepsilon_{_{MB}} \\
        
        0 & 0 & i\varepsilon_{_{MB}}^{*} & \omega  
    \end{matrix}
    \right)\,,
    \label{Gr-1}
\end{equation}
\end{widetext}
where $\Gamma_A^{\alpha}=2\pi|V_{\alpha}|^2\rho_{\alpha}$, $\Gamma_B^{\alpha}=2\pi|W_{\alpha}|^2\rho_{\alpha}$, and we have defined the function
\begin{equation}
    K = i \left[\sqrt{\Gamma_{A}^{S}\Gamma_{B}^{S}}e^{i\phi/2}+\sqrt{\Gamma_{B}^{D}\Gamma_{A}^{D}}e^{-i\phi/2}\right]\,. 
\end{equation}

The retarded GF satisfies $\mathbf{G}^{r}(\omega)=[\mathbf{G}^{a}(\omega)]^{\dagger}$. $\mathbf{\Gamma}^{D(S)}$ is the line-width function denoting the coupling between the MZMs and the lead$-\text{D(S)}$, and is given by

\begin{equation}
    \mathbf{\Gamma}^{\alpha}= \left(
    \begin{matrix}
        
         \Gamma^{\alpha}_{AA}  & 0                     & \Lambda^{\alpha}_{AB} & 0 \\
        0                     & 0  & 0                     & 0 \\
        \Lambda^{\alpha}_{BA} & 0                     & \Gamma^{\alpha}_{BB}  & 0 \\
        0                     & 0 & 0                     & 0 
    \end{matrix}
    \right)
    \label{12}\,,
\end{equation}
where we have defined $\Lambda^{\alpha}_{\beta\beta'}=\sqrt{\Gamma^{\alpha}_{\beta\beta'}\Gamma^{\alpha}_{\beta'\beta}}$, and $\Gamma^{\alpha}_{\beta\beta'}=2\pi t_{\alpha,{\bf k},\beta}(\varphi_{ \alpha \beta})[t_{\alpha,{\bf k},\beta'}(\varphi_{ \alpha \beta'})]^{*}\rho_{\alpha}$ is the tunnel-coupling strength, with $\rho_{\alpha}$ being the local density of states in the lead $\alpha$.

To examine the thermoelectric properties, we consider the system in the linear response regime, characterized by a temperature difference $\Delta T$ between the two leads. In this framework, the charge and heat currents, $I_{\text{charge}}$ and $I_{\text{heat}}$, can be expressed as functions of the potential difference $\Delta V$ as

\begin{equation}
    I_{\text{charge}}=-e^2L_0\Delta V + \dfrac{e}{T}L_1\Delta T\label{icharge}\,,
\end{equation}
\begin{equation}
    I_{\text{heat}}=eL_1\Delta V - \dfrac{1}{T}L_2\Delta T \label{iheat}\,,
\end{equation}
where the integrals $L_{n}$ are obtained from,

\begin{equation}
    L_n(\mu, T)= \dfrac{1}{h}\int_{0}^{\infty} \left(-\dfrac{\partial f(\omega,\mu)}{\partial \omega}\right)(\omega-\mu)^{n}\mathcal{T}(\omega)\,\text{d}\omega\label{}\,,
\end{equation}
where $f(\omega, \mu)=[\exp \{(\omega -\mu)/k_{\text{B}}T\}+1 ]^{-1}$ is the Fermi distribution function and $k_{\text{B}}$ the Boltzmann constant. The Seebeck coefficient $S$, also known as thermopower, describes the relationship between the temperature difference $\Delta T$ and the resulting potential difference $\Delta V$ induced when the charge current vanishes,

\begin{equation}
    S(\mu)=-\dfrac{\Delta V}{\Delta T}=-\dfrac{1}{e T}\dfrac{L_1}{L_0} \label{}\,.
\end{equation}
The electrical conductance $\mathcal{G}(\mu)$ is defined as the ratio of the charge current to the potential difference when the temperature difference $\Delta T$ is zero. Similarly, the thermal conductance $\kappa(\mu)$ is defined as the ratio of the heat current to the temperature gradient when the charge current is zero. Based on Eqs.\ (\ref{icharge}) and (\ref{iheat}), both conductances can be expressed as:

\begin{equation}
    \mathcal{G}(\mu) = -\frac{I_{\text{charge}}}{\Delta V} = e^2 L_0 \,,
\end{equation}

\begin{equation}
    \kappa (\mu) = -\frac{I_{\text{heat}}}{\Delta T} = \frac{1}{T} \left( L_2 - \frac{L_1^2}{L_0} \right)\label{kappamu}\,.
\end{equation}
Note that Eq.\ (\ref{kappamu}) accounts only for the electronic contribution to the thermal conductance, assuming that the phononic contribution is negligible in the low-temperature regime (a few kelvins) typical of these systems.

To quantify the efficiency of our MZM thermoelectric setups, we calculate the dimensionless figure of merit $ZT$, defined as

\begin{equation}
    ZT = \dfrac{S^2 \mathcal{G} T}{\kappa}\label{ZT}\,.
\end{equation}

An means to improve the ZT factor involves exceeding the constraints imposed by the Wiedemann-Franz (WF) law, which dictates the ratio $\kappa/\mathcal{G}T=\mathcal{L}_{0}\equiv constant$ across all systems, where $\mathcal{L}_{0}=(\pi^2/3)(k_B/e)^2$ represents the Lorenz number. Although macroscopic materials typically adhere to the WF law, nanostructured systems have demonstrated exceptional capability as thermoconverters, effectively transcending this restriction \cite{vineis2010nanostructured}. The four quantities defined above, $\mathcal{G}$, $\kappa$, $S$, and $ZT$, can be obtained using the Sommerfeld expansion in the integrals $L_n$, yielding the following:

\begin{equation}
    L_0=\dfrac{1}{h}\left[\mathcal{T}^{(0)}+\dfrac{\pi^2}{6}\mathcal{T}^{(2)}\xi^2+\dfrac{7\pi^4}{360}\mathcal{T}^{(4)}\xi^4+\mathcal{O}(\xi^6)\right]\,,\label{eq18}
\end{equation}

\begin{equation}
    L_1=\dfrac{1}{h}\left[\dfrac{\pi^2}{3}\mathcal{T}^{(1)}\xi^2+\dfrac{7\pi^4}{90}\mathcal{T}^{(3)}\xi^4+\mathcal{O}(\xi^6)\right]\,,\label{eq19}
\end{equation}

\begin{equation}
    L_2=\dfrac{1}{h}\left[\dfrac{\pi^2}{3}\mathcal{T}^{(0)}\xi^2+\dfrac{7\pi^4}{30}\mathcal{T}^{(2)}\xi^4+\mathcal{O}(\xi^6)\right]\,,\label{eq20}
\end{equation}
where $\mathcal{T}^{(n)}(\mu)=(d^n\mathcal{T}/d\omega^n)(\mu)$, and $\xi=k_BT$.

We also investigated the behavior of the spectral function, since it is closely related to resonances in the conductance. The spectral function is expressed as
\begin{equation}
    \text{A}(\omega)=-\dfrac{1}{\pi}\text{Im}\left[\text{Tr}\{\mathbf{G}^{r}(\omega)\}\right]\label{13}\,,
\end{equation}
and the spectral function for each TSCN are expressed as

\begin{equation}
    \text{A}_{jA}(\omega)=-\dfrac{1}{\pi}\text{Im}\{\mathbf{G}^{r}_{jj}(\omega)\}\,,
\end{equation}

\begin{equation}
    \text{A}_{jB}(\omega)=-\dfrac{1}{\pi}\text{Im}\{\mathbf{G}^{r}_{j+2, j+2}(\omega)\}\,,
\end{equation}
where $j=1,2$.
Moreover, the complete GF poles are closely related to the eigenvalues of the isolated TSCN-TSCN (disconnected from leads), and give reliable information about the energy localization of the system's states. The eigenvalues equation can be written as
\begin{equation}
    \omega^{4} -\varepsilon_{_{MA}}^2 \varepsilon_{_{MB}}^2 = 0\,.
\end{equation}
obtaining 2-degenerate solutions in the form

\begin{equation}
    \omega_{[\pm]}=\pm \sqrt{\varepsilon_{_{MA}} \varepsilon_{_{MB}}}\,.\label{eq25}
\end{equation}
For instance, in the particular case of long wire limit for both TSCN ($\varepsilon_{_{MA}}=\varepsilon_{_{MB}}=0$),

\begin{equation}
    \omega_{[\pm]}=0\,.
\end{equation}

\section{Results}\label{secIII}

\begin{figure}
    \centering
    \includegraphics[width=1.0\linewidth]{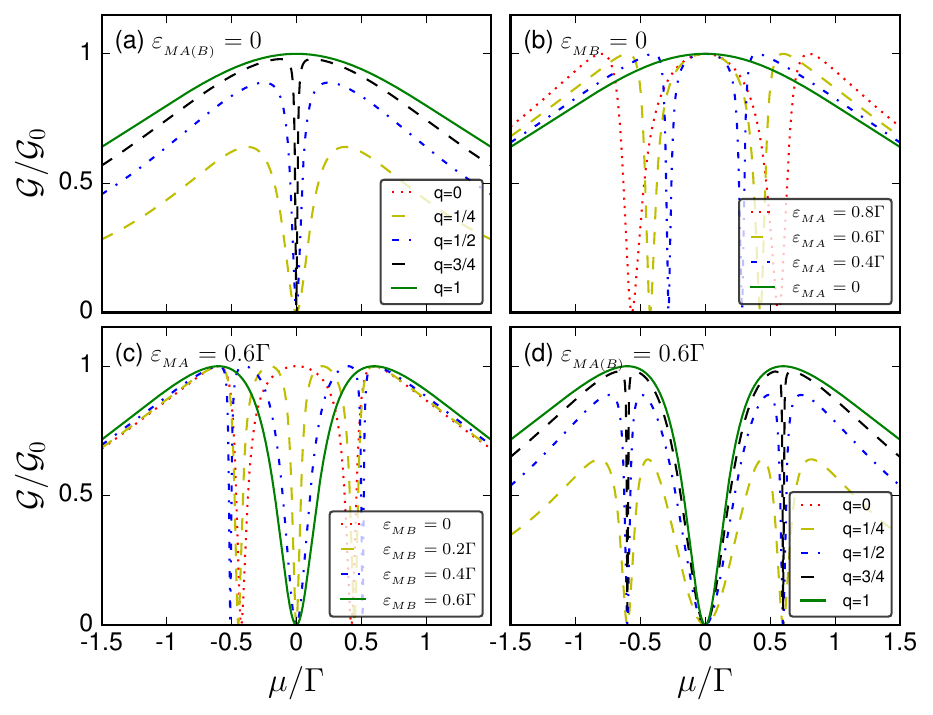}
    
    \caption{Electronic conductance $\mathcal{G}$ as a function of chemical potential $\mu$ without magnetic flux ($\phi=0$), we show the cases (a) $\varepsilon_{MA(B)}=0$ (long wire limit) with $q\in[0,1]$, (b) $\varepsilon_{_{MB}}=0$ with $\varepsilon_{_{MA}}/\Gamma=\{0, 0.4, 0.6, 0.8\}$, (c) $\varepsilon_{_{MA}}=0.6\Gamma$ with $\varepsilon_{_{MB}}/\Gamma=\{0,0.2, 0.4, 0.6\} $, and (d) $\varepsilon_{_{MA(B)}}=0.6\Gamma$ with $q\in[0,1]$. The dimensionless asymmetry parameter $q$ corresponds to a close(open) system for $q = 1$($q = 0$), respectively.}
    \label{fig2}
\end{figure}

We have considered the wide-band approximation, in which $\rho_{\alpha}$ has an approximately constant value, and then $\Gamma^{\alpha}_{B}$ and $\Gamma^{\alpha}_{B}$ are energy independent. Thus, we fixed the values $\Gamma^{S}_{A}=\Gamma^{D}_{A}=\Gamma^{S}_{B}=\Gamma^{D}_{B}=\Gamma$, and $q$ is a dimensionless parameter with $q=1(q=0)$ corresponding to a close(open) system. In the following, all energy parameters are given in units of $\Gamma$. In order to consider realistic parameters with experiments, the values of $\Gamma$ can be considered from a few to hundreds of meV. We assume a background temperature of $T = 1$\,K, well below typical superconductor critical temperatures \cite{nagamatsu2001:Nature}.

\begin{figure*}
    \centering
    \includegraphics[width=1.0\linewidth]{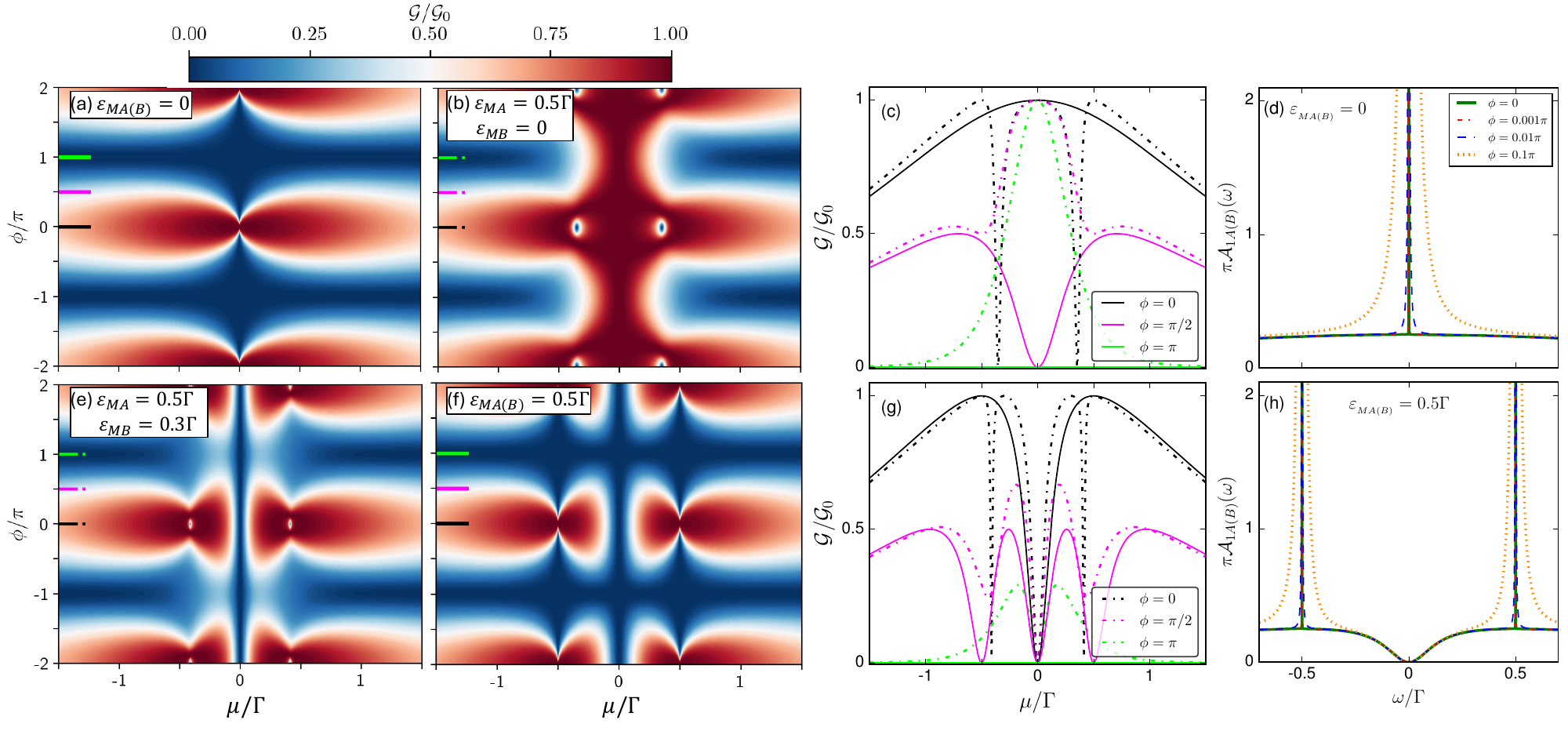}
    
    \caption{Colormap of the electronic conductance $\mathcal{G}$ as a function of both the magnetic flux $\phi$ and the chemical potential $\mu$, where red (blue) represents maximum (minimum) value for (a) $\varepsilon_{_{MA(B)}}=0$, (b) $\varepsilon_{_{MA}}=0.5\Gamma$ and $\varepsilon_{_{MB}}=0$, (e) $\varepsilon_{_{MA}}=0.5\Gamma$ and $\varepsilon_{_{MB}}=0.3\Gamma$, and (f) $\varepsilon_{_{MA(B)}}=0.5\Gamma$. In panel (c) and (g) we show the electronic conductance $\mathcal{G}$, as a function of the chemical potential $\mu$, for specific values of the magnetic flux $\phi=\{0, \pi/2, \pi\}$ in black, magenta, and light green color, respectively. The solid (dash-dotted) line correspond to symmetrical (asymmetrical) configuration of the length of both TSCN, where $\varepsilon_{_{MA(B)}}=0$ ($\varepsilon_{_{MA(B)}}=0.5\Gamma$) and $\varepsilon_{_{MA}}=0.5\Gamma$ with $\varepsilon_{_{MB}}=0$ ( $\varepsilon_{_{MA}}=0.5\Gamma$ with $\varepsilon_{_{MB}}=0.3\Gamma$), are shown in the top (bottom) panel. The spectral function $\mathcal{A}_{1A(B)}$ as a function of energy $\omega$ is shown in the panel (d) $\varepsilon_{_{MA(B)}}=0$ and (h) $\varepsilon_{_{MA(B)}}=0.5\Gamma$. We use specific values of the magnetic flux $\phi=\{0, 0.001\pi, 0.01\pi,0.1\pi\}$ in solid green, dash-doted red, dashed blue, and dotted orange line, respectively.}
    \label{fig3}
\end{figure*}

First, we analyze the electronic transport in the system by considering three scenarios: only one, both, or neither TSCN in the long-wire limit. In this regime, the MZMs in each TSCN can interfere in the transmission process depending on whether they are coupled ($\varepsilon_{_{M\beta}}\neq0$) or decoupled ($\varepsilon_{_{M\beta}}=0$). The electronic conductance $\mathcal{G}$ as a function of the chemical potential $\mu$ is shown in Fig.~\ref{fig2}. Panel (a) corresponds to the case where both TSCNs are in the long wire limit ($\varepsilon_{_{MA(B)}}=0$), ensuring that the MZMs are decoupled from the external ends of each TSCN. The dimensionless asymmetry parameter $q$ characterizes the openness of the system: $q = 1$ corresponds to a closed system, while $q = 0$ corresponds to an open one.
We obtain a Breit--Wigner resonance centered at $\mu = 0$ for $q = 1$, while an antiresonance at $\mu = 0$ appears for $q \neq 1$. We also show that the electronic conductance progressively decreases to zero as the circuit transitions from a closed system ($q = 1$, solid green line) to an open system ($q = 0$, dotted red line). This behavior arises because the transmission coefficient is proportional to the line width function $\Gamma^{\alpha}(q)$; that is, as $q$ tends to zero, the matrix elements connecting the leads and the TSCNs also tend to zero. Panel (b) corresponds to a closed system ($q = 1$), where one TSCN is in the long wire limit ($\varepsilon_{_{MB}} = 0$), while in the other TSCN, the MZM coupling $\varepsilon_{_{MA}} \geq 0$ is varied.
This results in a Breit--Wigner resonance centered on $\mu = 0$, reaching the value $\mathcal{G} = e^2/h$. When $\varepsilon_{_{MA}} \neq 0$, the electronic conductance consists of a central Breit--Wigner resonance and two lateral antiresonances located at $\mu = \pm \varepsilon_{_{MA}}/\sqrt{2}$. In panel (c), we fix the MZM coupling $\varepsilon_{_{MA}} = 0.6\Gamma$ and vary the coupling $\varepsilon_{_{MB}}$. The electronic conductance exhibits an antiresonance at $\mu = 0$ and two lateral antiresonances at energies $\mu = \pm (\varepsilon_{_{MA}} + \varepsilon_{_{MB}})/2$ when $\varepsilon_{_{MB}} \neq 0$, except in the symmetric case $\varepsilon_{_{MA(B)}} = 0.6\Gamma$ (solid green line), where the two lateral antiresonances evolve into resonances at $\mu = \pm 0.6\Gamma$. In panel (d), we show that the symmetric MZM-coupling configuration ($\varepsilon_{_{MA(B)}} = 0.6\Gamma$), shown as the solid green line, gives rise to antiresonances at $\mu = \pm 0.6\Gamma$ when $q \neq 1$. We note that the position of these antiresonances is independent of the value of $q$, as they are centered at the system’s eigenenergies, which are determined independently of $q$, as can be seen in Eq.~(\ref{eq25}).

We study the electronic transport in a closed system ($q = 1$) in the presence of a magnetic flux across the interferometer (i.e., $\phi \neq 0$). Figure~\ref{fig3} shows a colormap of the electronic conductance $\mathcal{G}$ as a function of the dimensionless magnetic flux $\phi$ and $\mu$. In FIG.~\ref{fig3}(a), we consider the long-wire limit for both TSCNs ($\varepsilon_{_{MA(B)}} = 0$). At zero energy, the electronic conductance is $\mathcal{G} = e^2/h$ for $\phi = 2n\pi$ ($n \in \mathbf{Z}$), and drops to zero for $\phi \neq 2n\pi$, where the magnetic flux induces transport suppression over a wide range of values, reaching total reflection at $\phi = (2n - 1)\pi$. Figure~\ref{fig3}(b) shows the case with $\varepsilon_{_{MA}} = 0.5\Gamma$ and $\varepsilon_{_{MB}} = 0$. At zero energy, the electronic conductance remains $\mathcal{G} = e^2/h$ and is invariant under symmetry-breaking induced by the magnetic flux. Figure~\ref{fig3}(e) shows $\mathcal{G}$ as a function of $\mu$ and $\phi$ for the case where both TSCNs are outside the long-wire limit, but with different lengths, i.e., $\varepsilon_{_{MA}} = 0.5\Gamma$ and $\varepsilon_{_{MB}} = 0.3\Gamma$.
We observe that, regardless of the magnetic flux, the linear conductance exhibits an antiresonance at zero energy. For the particular case $\phi = 2n\pi$, two lateral antiresonances appear at $\mu = \pm (\varepsilon_{_{MA}} + \varepsilon_{_{MB}})/2$. Figure~\ref{fig3}(f) shows the electronic conductance $\mathcal{G}$ for the case where both TSCNs are outside the long-wire limit, i.e., $\varepsilon_{_{MA}} = \varepsilon_{_{MB}} = 0.5\Gamma$. Again, the linear conductance displays an antiresonance at zero energy, independent of the magnetic flux. The suppression of transport as a function of $\mu$ is recovered—similar to the behavior in Fig.~\ref{fig3}(a)—for values $\phi = (2n - 1)\pi$. This behavior appears only for symmetric configurations of the MZM couplings, i.e., when $\varepsilon_{_{MA}} = \varepsilon_{_{MB}}$. Figures~\ref{fig3}(c) and \ref{fig3}(g) show the electronic conductance $\mathcal{G}$ as a function of $\mu$ for fixed values of the magnetic flux: $\phi = 0$, $\phi = \pi/2$, and $\phi = \pi$, represented by black, magenta, and light green lines, respectively. The solid (dash-dotted) lines correspond to symmetric (asymmetric) configurations of the MZMs couplings.
We observe the phenomenon of total reflection ($\mathcal{G} = 0$) in the symmetric case $\varepsilon_{_{MA}} = \varepsilon_{_{MB}}$ for $\phi = \pi$ (solid light green line in both panels), which is an energy-independent behavior. Figures~\ref{fig3}(d) and \ref{fig3}(h) display the spectral functions $\mathcal{A}_{1\beta}$ as a function of the energy $\omega$ for magnetic flux values $\phi = 0$, $\pi/1000$, $\pi/100$, and $\pi/10$, shown in green (solid line), red (dash-dotted line), blue (dashed line), and orange (dotted line), respectively. For a symmetric configuration of the MZMs coupling ($\varepsilon_{_{MA}} = \varepsilon_{_{MB}}$), we find $\mathcal{A}_{1A} = \mathcal{A}_{1B}$, with parameters [panel (d)] $\varepsilon_{_{MA(B)}} = 0$ and [panel (h)] $\varepsilon_{_{MA(B)}} = 0.5\Gamma$. In the long-wire limit ($\varepsilon_{_{MA(B)}} = 0$), shown in Fig.~\ref{fig3}(d), we observe a zero-width resonance localized at $\omega = 0$ for $\phi = 0$ (solid green line), corresponding to a true BIC, since these states do not contribute to the electronic conductance $\mathcal{G}$. These states acquire a finite width as the magnetic flux increases ($\phi \neq 0$), becoming quasi-BICs and contributing to the transmission in the form of antiresonances.
In Fig.~\ref{fig3}(h), we consider the case where both TSCNs have finite and equal lengths ($\varepsilon_{_{MA(B)}} = 0.5\Gamma$). For $\phi = 0$, we obtain two symmetric lateral BICs located at $\omega = \pm \varepsilon_{_{MA(B)}} = \pm 0.5\Gamma$, in agreement with Eq.~(\ref{eq25}). These states do not have projections in the electronic conductance, as shown in Fig.~\ref{fig3}(g) for $\phi = 0$. When $\phi \neq 0$, the two symmetric lateral BICs in the spectral function acquire a finite width, thus becoming quasi-BICs.

\begin{figure}[t]
    \centering
    \includegraphics[width=1.0\linewidth]{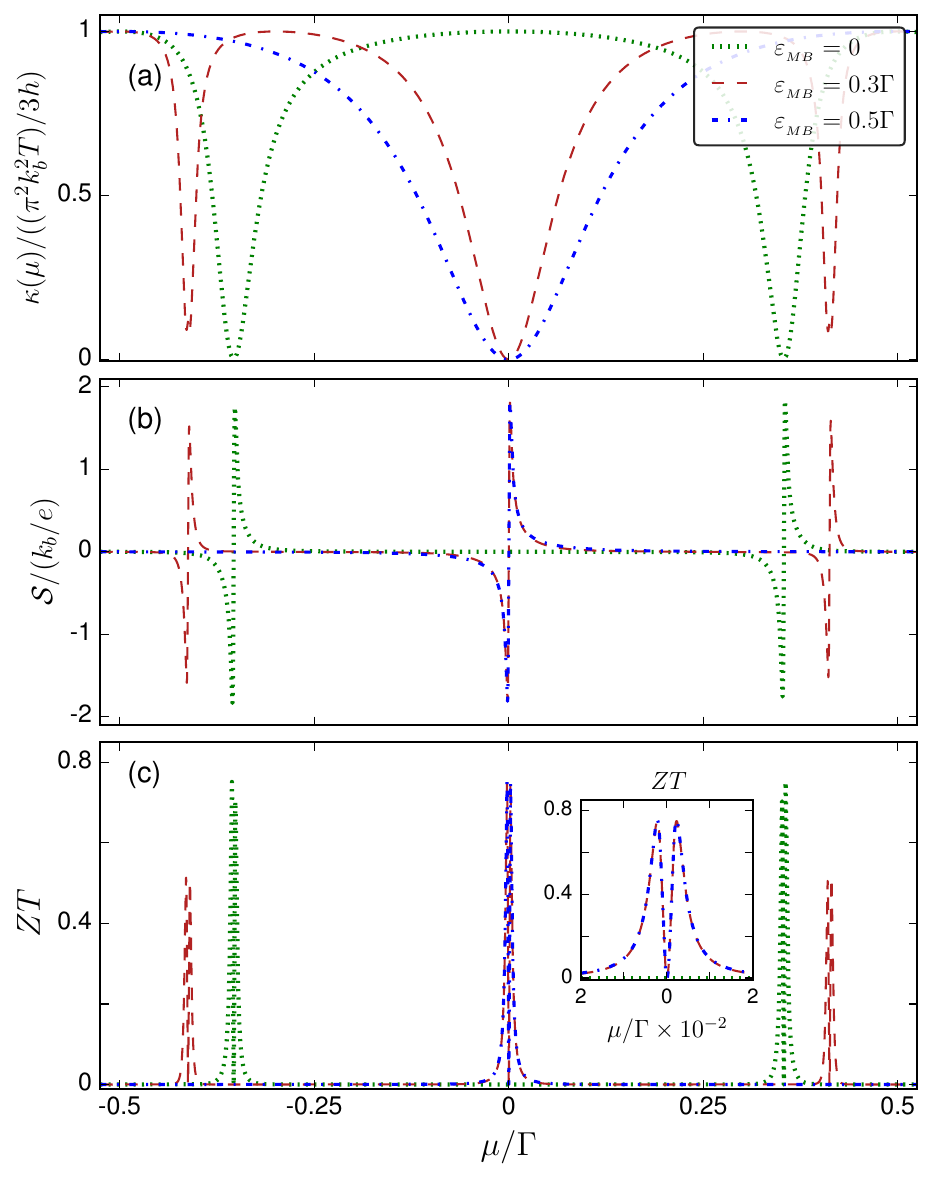}
    \caption{Thermal quantities as a function of chemical potential $\mu$. We show the (a) thermal  conductance $\kappa$, (b) Seebeck coefficient $S$, and (c) figure of merit $ZT$, for the cases $\varepsilon_{_{MB}}=0$ (dotted green line), $\varepsilon_{_{MB}}=0.3\Gamma$ (dashed red line), and $\varepsilon_{_{MB}}=0.5\Gamma$ (dash-dotted blue line), with $\varepsilon_{_{MA}}=0.5\Gamma$ and $\phi=0$ for all panels.}\label{fig4}
\end{figure}

\begin{figure}[t]
    \centering
    \includegraphics[width=1.0\linewidth]{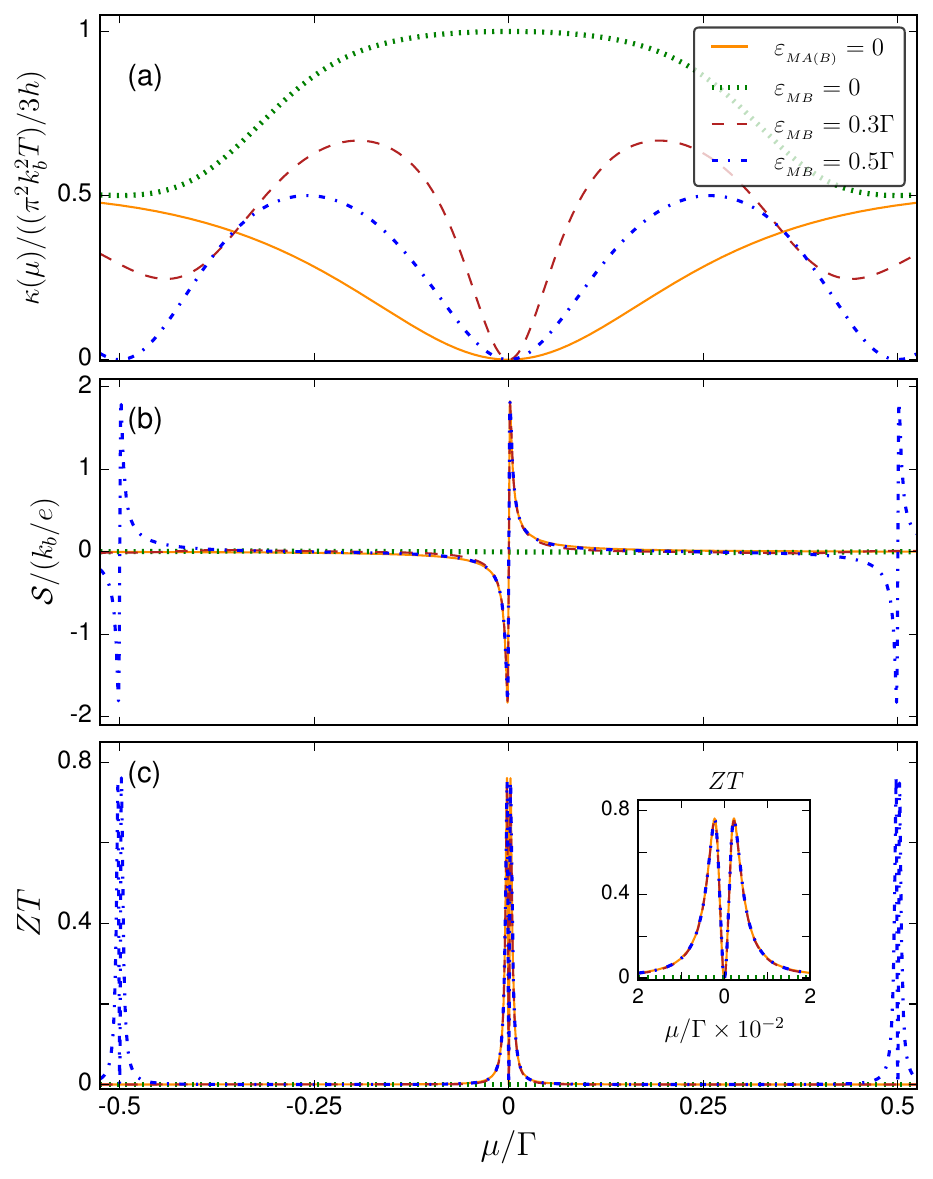}
    \caption{Thermal quantities as a function of chemical potential $\mu$. We show the (a) thermal  conductance $\kappa$, (b) Seebeck coefficient $S$, and (c) figure of merit $ZT$, for the cases $\varepsilon_{_{MA(B)}}=0$ (solid orange line),  $\varepsilon_{_{MB}}=0$ (dotted green line), $\varepsilon_{_{MB}}=0.3\Gamma$ (dashed red line), and $\varepsilon_{_{MB}}=0.5\Gamma$ (dash-dotted blue line), with $\varepsilon_{_{MA}}=0.5\Gamma$ for (b)-(d), and magnetic flux $\phi=\pi/2$ in all panels.}\label{fig5}
\end{figure}

We now focus our attention on the thermoelectric properties of the system.
Figure~\ref{fig4} shows the [panel (a)] thermal conductance $\kappa$, [panel (b)] Seebeck coefficient $S$, and [panel (c)] figure of merit $ZT$ as functions of $\mu$, in the absence of magnetic flux ($\phi = 0$). We fix $\varepsilon_{_{MA}} = 0.5\Gamma$, and the second TSCN takes the values $\varepsilon_{_{MB}} = 0$ (dotted green line), $\varepsilon_{_{MB}} = 0.3\Gamma$ (dashed red line), and $\varepsilon_{_{MB}} = 0.5\Gamma$ (dash-dotted blue line). The thermal conductance in Fig.~\ref{fig4}(a) exhibits a behavior similar to that of the electronic conductance (see, for instance, Fig.~\ref{fig3}), where resonances and antiresonances depend on the MZM couplings of each TSCN. The cases with $\varepsilon_{_{MB}} = 0$, $\varepsilon_{_{MB}} = 0.3\Gamma$, and $\varepsilon_{_{MB}} = 0.5\Gamma$ correspond to the dash-dotted black line in Fig.~\ref{fig3}(c), the dash-dotted black line in Fig.~\ref{fig3}(g), and the solid black line in Fig.~\ref{fig3}(g), respectively. The Seebeck coefficient is shown in Fig.~\ref{fig4}(b), and is an odd function of $\mu$. The changes in $S$, from minimum to maximum, are centered at $\mu = \pm \varepsilon_{_{MA}}/\sqrt{2}$ (dotted green line), $\mu = 0$ and $\mu = \pm (\varepsilon_{_{MA}} + \varepsilon_{_{MB}})/2$ (dashed red line), and $\mu = 0$ (dash-dotted blue line), which coincide with the positions of antiresonances in the thermal conductance shown in Fig.~\ref{fig4}(a). The thermoelectric efficiency is characterized by the extrema of the figure of merit, $ZT$, shown in Fig.~\ref{fig4}(c). We observe that the maxima of $ZT$ appear in pairs and are centered at the same energies as those found in the thermopower and thermal conductance. In the inset, we show a zoomed view where the maxima of $ZT$ exhibit a symmetric behavior centered at $\mu = 0$. At this point, we can express that the thermoelectric properties of the system are strongly influenced by the coupling between the MZMs in each TSCN. The location of resonances and antiresonances in the electronic and thermal conductances correlates with features in the Seebeck coefficient and thermoelectric efficiency. In particular, symmetric configurations of MZM couplings lead to well-defined antiresonances and enhanced thermoelectric response.

Figure~\ref{fig5} shows the [panel (a)] thermal conductance $\kappa$, [panel (b)] Seebeck coefficient $S$, and [panel (c)] figure of merit $ZT$ as functions of $\mu$, in the presence of magnetic flux ($\phi = \pi/2$). We first present the case $\varepsilon_{_{MA(B)}} = 0$ (solid orange line). Then, we fix $\varepsilon_{_{MA}} = 0.5\Gamma$ and vary the second TSCN coupling as $\varepsilon_{_{MB}} = 0$ (dotted green line), $\varepsilon_{_{MB}} = 0.3\Gamma$ (dashed red line), and $\varepsilon_{_{MB}} = 0.5\Gamma$ (dash-dotted blue line). The thermal conductance in Fig.~\ref{fig5}(a) exhibits behavior similar and proportional to that of the electronic conductance. We observe this correspondence in Figs.~\ref{fig3}(c) and \ref{fig3}(g), where the solid and dash-dotted magenta lines represent symmetric and asymmetric MZM-coupling configurations, respectively. As before, the positions of resonances and antiresonances depend on the MZMs couplings of each TSCN.
The Seebeck coefficient is shown in Fig.~\ref{fig5}(b), and is an odd function of the chemical potential $\mu$. The variations in $S$, from minimum to maximum, are centered at $\mu = \pm \varepsilon_{_{MA(B)}}$ and $\mu = 0$ (dash-dotted blue line), and at $\mu = 0$ for the solid orange and dashed red lines. These positions coincide with the locations of antiresonances in the thermal conductance shown in Fig.~\ref{fig5}(a). The maxima of $ZT$ appear in pairs [Fig.~\ref{fig5}(c)], and are centered at the same energies as those observed in the thermopower and thermal conductance. In the inset, a zoomed view reveals that the maxima of $ZT$ exhibit a symmetric behavior centered at $\mu = 0$. We observe that the BICs present in the system for $\phi = 0$ do not affect the thermoelectric quantities. However, when $\phi \neq 0$, these BICs are destroyed and become quasi-BICs, which enhance the thermoelectric efficiency. This effect is particularly evident in Fig.~\ref{fig5}(c) for the cases $\varepsilon_{_{MA(B)}} = 0$ (solid orange line) and $\varepsilon_{_{MA(B)}} = 0.5\Gamma$ (dash-dotted blue line).

\begin{figure}[t]
    \centering
    \includegraphics[width=1.0\linewidth]{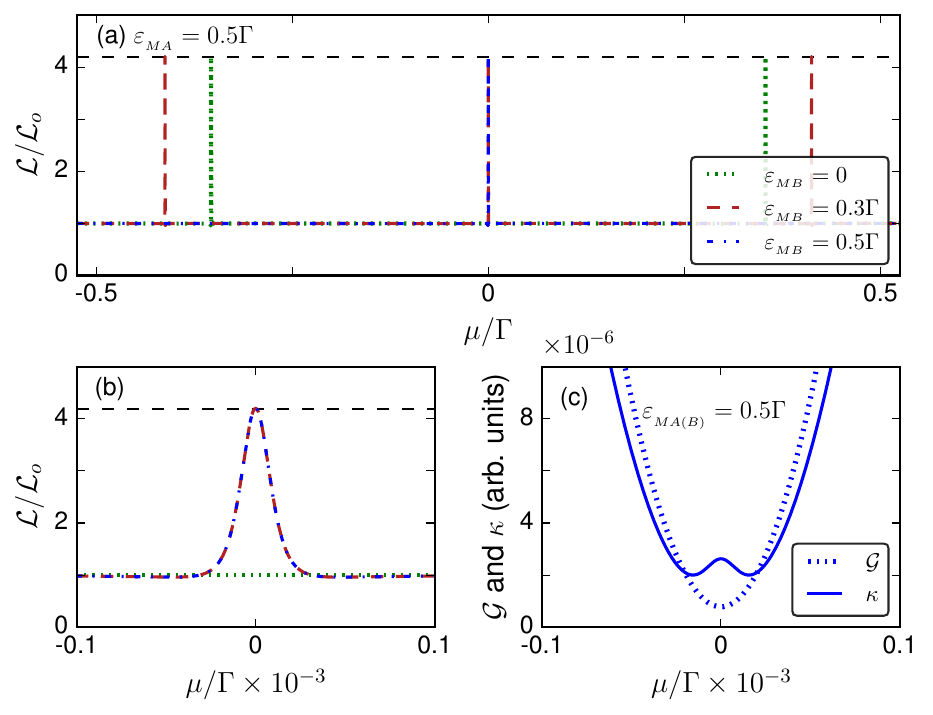}
    \caption{We show (a) Lorenz ratio $\mathcal{L}$ as a function of $\mu$, for the cases $\varepsilon_{_{MB}}=0$ (dotted green line), $\varepsilon_{_{MB}}=0.3\Gamma$ (dashed red line), and $\varepsilon_{_{MB}}=0.5\Gamma$ (dash-dotted blue line), with $\varepsilon_{_{MA}}=0.5\Gamma$ and $\phi=0$ for all panels. The horizontal dashed black line corresponds to the universal maximum value of 4.19$\mathcal{L}_0$. In panel (b) we show a zoom of panel (a) centered at $\mu=0$. In panel (c) we plot both conductances ($\mathcal{G}$ and $\kappa$, in arbitrary units) for the case $\varepsilon_{_{MA(B)}}=0.5\Gamma$, in the same energy range that in panel (b). }\label{fig6}
\end{figure}

In Fig.~\ref{fig6} we study the fulfillment of the WF law by plotting the Lorenz ratio $\mathcal{L}$ in units of the Lorenz number $\mathcal{L}_0$, as a function of chemical potential $\mu$. The panel \ref{fig6}(a) shows the cases $\varepsilon_{_{MB}}=0$ (dotted green line), $\varepsilon_{_{MB}}=0.3\Gamma$ (dashed red line) and $\varepsilon_{_{MB}}=0.5\Gamma$ (dashed-dotted blue line), with $\varepsilon_{_{MA}}=0.5\Gamma$ and $\phi=0$. The Lorenz ratio $\mathcal{L}=\kappa/\mathcal{G}T=\mathcal{L}_{0}$ for almost all values of $\mu$, however $\mathcal{L}$  deviates from $\mathcal{L}_{0}$ around $\mu=\{0, \pm\varepsilon_{_{MA}}/\sqrt{2},\pm(\varepsilon_{_{MA}}+\varepsilon_{_{MB}})/2 \}$, where $\mathcal{L}$ reaches the maximum $\mathcal{L}_{max}=4.19\mathcal{L}_{0}$. Fig.~\ref{fig6}(b) is a zoom of Fig.~\ref{fig6}(a) centered at $\mu=0$. We can observe from Eq. (\ref{eq19}), that the expansion for the integral $L_1$  contains only odd derivatives of the transmission $\mathcal{T}$, which is dominated by the term proportional to $\mathcal{T}^{(1)}$, which vanishes at the antiresonance energy. As a result of this, the thermal conductance has a small peak in the antiresonance region due to the term $L_1^2/L_0$, in Eq. (\ref{kappamu}), it falls to zero, while the electronic conductance $\mathcal{G}$ presents a single minimum, as can be seen in panel \ref{fig6}(c). Both curves present different shapes in a small region around the antiresonance energy, which results in the violation of the WF law.
\section{Summary}\label{secIV}

We studied a system composed of two normal leads coupled to two TSCNs, each hosting MZMs at their ends, arranged in an interferometer configuration. We focused on the electronic and thermal conductances between the leads, as well as on the spectral functions of the MZMs and thermoelectric quantities. The latter were obtained using the GF formalism, while thermoelectric properties were calculated via the Sommerfeld expansion. We reported the phenomenon of total reflection at magnetic flux values $\phi = (2n - 1)\pi$ for symmetric MZM-coupling configurations, that is, when both TSCNs have the same length.
In addition, for magnetic flux values $\phi = 2n\pi$, we identified the formation of BICs, characterized by zero-width resonances in the spectral functions. These states also emerge under symmetric MZM coupling and behave as ghost Fano-Majorana anomalies, since they do not contribute to the electronic conductance. We also found that these BICs are destroyed as the magnetic flux deviates from $\phi = 2n\pi$. For $\phi \neq 0$, the BICs acquire a finite width, becoming quasi-BICs, and manifest themselves as antiresonances in both electronic and thermal conductances at the same characteristic energies. These results demonstrate that BICs in the system can be controlled via the external magnetic flux, and their transformation into quasi-BICs leads to enhancements in thermopower $S$ and thermoelectric figure of merit $ZT$, by means of a violation of the WF law.

\begin{acknowledgments}
A.P.G. is grateful for the funding of scholarship ANID-Chile No. 21210410 and FONDECYT grant 1201876. D.Z. acknowledges support from USM-Chile under Grant PI-LIR-2022-13. J.P.R.-A is grateful for the financial support of FONDECYT Iniciaci\'on grant No. 11240637. P.A.O. acknowledges support from FONDECYT grants 1201876 and 1220700.  
\end{acknowledgments}

\section*{DATA AVAILABILITY STATEMENT}

Data will be made available on reasonable request.

\section*{CONFLICTS OF INTEREST}

The authors decl are that they have no conflict of interest.

\section*{AUTHOR CONTRIBUTION STATEMENT}

All authors contributed equally and significantly in writing this article. All authors read and approved the final manuscript.

\onecolumngrid
\appendix
\section{Green function}\label{AppendixA}
The full Green function is obtained from Eq. (\ref{Gr-1}), in the form
\begin{equation}
    \mathbf{G}^{r}= \dfrac{1}{D}\left( 
    \begin{matrix}
        G_{11} & G_{12} & -K\omega^{2} & -iK\varepsilon_{_{MB}}\omega\\ 
        
        G_{21} & G_{22}  & iK\varepsilon_{_{MA}}\omega & -K\varepsilon_{_{MA}}\varepsilon_{_{MB}} \\
        
        -K\omega^{2} & -iK\varepsilon_{_{MA}}\omega & G_{33}  & G_{34} \\
        
        iK\varepsilon_{_{MB}}\omega & -K\varepsilon_{_{MA}}\varepsilon_{_{MB}} & G_{43} & G_{44}  
    \end{matrix}
    \right)\,,
    \label{Gr}
\end{equation}

with the matricial elements

\begin{equation}
    G_{11(33)}=-\omega\left[\varepsilon_{_{MB(A)}}^{2}-i\left[\sum_{\alpha}\Gamma^{\alpha}_{B(A)}\right]\omega-\omega^{2}\right]\,,
\end{equation}
\begin{equation}
    G_{22(44)}=-\varepsilon_{_{MB(A)}}^{2}\left[i\sum_{\alpha}\Gamma^{\alpha}_{A(B)}+\omega\right]-\omega K^{2}+\omega\left[i\sum_{\alpha}\Gamma^{\alpha}_{A}+\omega\right]\left[i\sum_{\alpha}\Gamma^{\alpha}_{B}+\omega\right]\,,
\end{equation}
\begin{equation}
    G_{12}=-G_{21}=-i\varepsilon_{_{MA}}\left[\varepsilon_{_{MB}}^{2}-i\sum_{\alpha}\Gamma^{\alpha}_{B}\omega-\omega^{2}\right]\,,
\end{equation}
\begin{equation}
    G_{34}=-G_{43}=-i\varepsilon_{_{MB}}\left[\varepsilon_{_{MA}}^{2}-i\sum_{\alpha}\Gamma^{\alpha}_{A}\omega-\omega^{2}\right]\,,
\end{equation}

and the denominator $D$,

\begin{equation}
     D = \left[\varepsilon_{_{MA}}^{2}-i\sum_{\alpha}\Gamma^{\alpha}_{A}+\omega\right]\left[\varepsilon_{_{MB}}^{2}-i\sum_{\alpha}\Gamma^{\alpha}_{B}+\omega\right]- K^{2} \omega^{2}\,.
\end{equation}

\twocolumngrid


\bibliographystyle{apsrev4-1}
\bibliography{bibprepint}

\end{document}